# High capacity $H_2$ adsorption over $Si_4Li_n$ (n=1-3) binary clusters: A DFT study

Ankita Jaiswal[1,a], Sridhar Sahu [1,b] *

[1]*Computational Materials Research Lab, Dept. of Physics, Indian Institute of Technology (Indian School of Mines) Dhanbad, Jharkhand, 826004, India*

[a]ankitajaiswal.iitism@gmail.com, [b]sridharsahu@iitism.ac.in

**Abstract**

This paper presents a detailed study of the hydrogen adsorption properties of small silicon-lithium binary nanoclusters. The stabilities of $H_2$ adsorbed binary clusters are assured by maximum hardness and minimum electrophilicity principle. Detail computational studies demonstrate that each Li in $Si_4Li$, $Si_4Li_2$ and $Si_4Li_3$ binary clusters can adsorb a maximum of $5H_2$, $4H_2$ and $3H_2$ molecules respectively, resulting in a total gravimetric density of 7.8%, 11.3% and 12%. The adsorption energy is observed to be in the range of 0.12eV-0.17eV, indicating that the adsorption of $H_2$ molecules occurs in a quasi-molecular fashion via physisorption. Hence, our study concludes that the studied silicon -lithium binary clusters can be considered as potential candidates for hydrogen storage systems.

**Introduction**

Globalization, industrialization, increased population, and demand for sustainable development has heightened our energy need to a large extent. As a result, the consumption of prime sources of energy, i.e., fossil fuels, has also increased at an alarming rate [1]. Besides, greenhouse gases produced from fossil fuel combustion are the leading cause of global warming. In order to keep global warming and climatic disturbances in check, we have to search for some clean sources of energy that can fulfill our energy needs and prevent our environment from further adverse effects. This scenario has heightened the urgent search for clean energy sources. Literature survey reveals that hydrogen can be regarded as one of the potential candidates for clean energy sources because of its abundance, maximum energy density, and carbon-free combustion [2]. However, the production and storage of hydrogen is the more significant challenge that has prohibited the development of the hydrogen-based economy.

Studies also report that conventional hydrogen storage methods like storing gaseous hydrogen in cylinders at high pressure (>700 bar) and storing liquid hydrogen at very low temperature (~20K) are quite costly and can cause safety issues during transportation [3]. According to the United States-Department of energy (US-DOE), a material-based template can be counted as a potential candidate for hydrogen storage only if it can achieve a gravimetric density greater than 5.5 wt% and can undergo desorption between the temperature range of -40°C to 85°C [4]. Therefore, extensive research is still going on to search and design storage materials that can fulfill US-DOE requirements and be used efficiently for practical applications. For this purpose, the $H_2$ storage in different materials like metal hydrides [5], MOF [6], graphene [7] etc., have been studied thoroughly. In recent years, hydrogen storage in nanoclusters [8], fullerenes [9], and binary clusters [10], has attracted much popularity.

In this paper, we have addressed the hydrogen storage capacities of small silicon-lithium binary clusters $Si_4Li$, $Si_4Li_2$, and $Si_4Li_3$, the structural and stability of which are already studied by E. Osorio et al. [11].

**Computational and theoretical details**

The DFT calculations are computed using m06/6-311+g(d,p) using *Gaussian09* software [12, 13]. Topological analysis and bonding properties of both the host and the sequential $H_2$ adsorbed complexes are studied using the concept of Bader's Quantum Theory of Atoms In Molecules (QTAIM) [14].

The stability of host and hydrogen adsorbed clusters are determined using conceptual density functional theory [15] and their corresponding HOMO-LUMO gap analysis. Reactivity descriptors like hardness ($\eta$), electrophilicity ($\omega$), and electronegativity ($\chi$) are calculated using the following relations-

$$\eta = \frac{(I-A)}{2} \quad (1)$$

$$\omega = \frac{\chi^2}{2\eta} \quad (2)$$

$$\chi = \frac{(I+A)}{2} \quad (3)$$

where I and A are the vertical ionization energy and electron affinity of the clusters calculated using Koopman's theorem [16]. The following relation can give the adsorption energy of the adsorbed $H_2$ molecules over the host clusters

$$E_{ads} = \frac{\{E_{Host} + nE_{H_2}\} - E_{complex}}{n} \quad (4)$$

Here $E_{Host}$, $E_{H_2}$ and $E_{complex}$ are the global minimum energy of the host, adsorbed $H_2$ molecules, and the hydrogen adsorbed complexes respectively.

**Results and discussions**

*Structure and stability analysis*

The host binary clusters $Si_4Li$, $Si_4Li_2$, and $Si_4Li_3$ and their sequential hydrogenated systems are optimized using the basis set and functional mentioned in the computational details to get the minimum energy structures. Vibrational analysis indicates the absence of imaginary frequencies, confirming that these binary clusters are stable. It is found that the optimized geometry of $Si_4Li$, $Si_4Li_2$ and $Si_4Li_3$ have $C_s$, $C_1$ and $C_s$ symmetry which is well in agreement with that reported by Osorio et al. The optimized geometries of both hosts and their saturated hydrogen adsorbed clusters are depicted in Fig. 1.

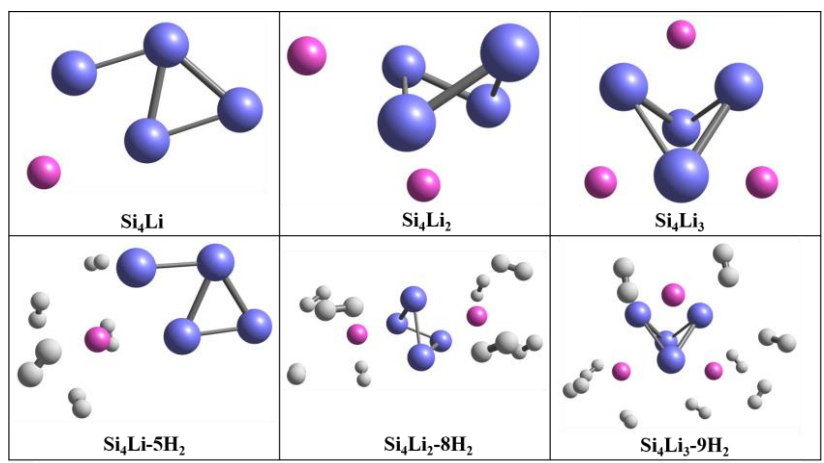

**Fig. 1**: Optimized geometries of bare hosts and their saturated $H_2$ adsorbed clusters.

Reactivity descriptors hardness ($\eta$), electrophilicity ($\omega$) and electronegativity ($\chi$) of the bare hosts and their saturated hydrogenated clusters are calculated using expressions (1), (2) and (3) respectively and is presented in Table 1. From the tabulated data, it is found that in all cases the hardness of the hydrogenated systems has increased as compared to the bare hosts whereas electrophilicity follows the reverse trend. This confirms that the maximum hardness and minimum electrophilicity principle is strictly followed, demonstrating that the hydrogenated systems are more stable than the bare hosts.

**Table 1**: Hardness ($\eta$), electrophilicity ($\omega$) and electronegativity ($\chi$) of hosts and saturated hydrogen adsorbed clusters.

| Cluster | $\eta$ (eV) | $\omega$ (eV) | $\chi$ (eV) | Gravimetric density (wt%) |
|---|---|---|---|---|
| $Si_4Li$ | 1.103 | 5.076 | 3.347 | - |
| $Si_4Li$-$5H_2$ | 1.294 | 3.635 | 3.067 | 7.8 |
| $Si_4Li_2$ | 1.408 | 4.291 | 3.477 | - |
| $Si_4Li_2$-$8H_2$ | 1.635 | 3.16 | 3.214 | 11.3 |
| $Si_4Li_3$ | 1.416 | 3.184 | 3.003 | - |
| $Si_4Li_3$-$9H_2$ | 1.523 | 2.298 | 2.646 | 12 |

*$H_2$ adsorption over silicon-lithium binary clusters*

$H_2$ molecules are added sequentially over the binary clusters to get a detailed insight into the hydrogen adsorption over the studied $Si_4Li$, $Si_4Li_2$, and $Si_4Li_3$ binary clusters. Analysis reveals that each Li present in $Si_4Li$, $Si_4Li_2$, and $Si_4Li_3$ binary clusters can hold up to $5H_2$, $4H_2$ and $3H_2$ molecules, thus making the final hydrogenated systems as $Si_4Li$-$5H_2$, $Si_4Li_2$-$8H_2$ and $Si_4Li_3$-$9H_2$ respectively. The gravimetric density for $Si_4Li$, $Si_4Li_2$ and $Si_4Li_3$ binary clusters are thus found to be 7.8%, 11.3% and 12% respectively. In addition, the adsorption energy per $H_2$ molecule of the saturated hydrogen adsorbed $Si_4Li$, $Si_4Li_2$ and $Si_4Li_3$ binary clusters is found to be in the range of 0.12eV-0.17eV, which confirm that the $H_2$ molecules are adsorbed over the host in quasi-molecular fashion via Niu-Rao-Jena type of interaction [17]. Besides, the adsorption energy range demonstrates that these hydrogenated systems can be easily reversed at near ambient conditions. With the increase in the number of adsorbed, the average adsorption energy tends to decrease because of the steric hindrance and mutual repulsion between the nearby $H_2$ molecules. The variation of adsorption energy with sequential $H_2$ adsorption is depicted in the Fig. 2.

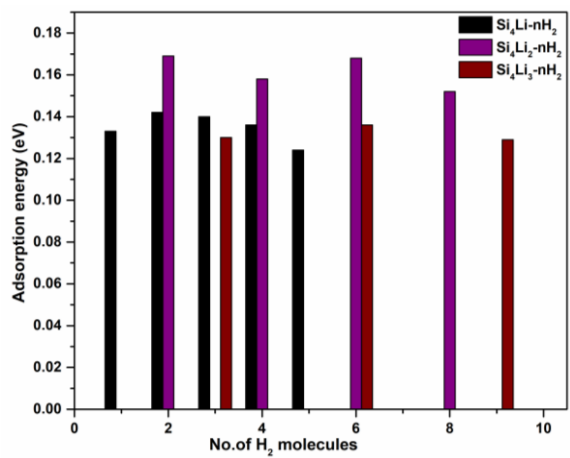

**Fig. 2**: Adsorption energy for sequential hydrogen adsorption over $Si_4Li$, $Si_4Li_2$ and $Si_4Li_3$ clusters.

*3.2.1 ESP maps and Hirshfeld charge analysis*

Electrostatic potential maps of bare silicon-lithium binary clusters and their sequential hydrogenated complexes are plotted to understand the variation of electron density and charge transfer as a result of sequential adsorption. It is plotted by surface mapping of the electrostatic potential on total electron density. In ESP maps, the electron density variation is given in terms of various colour codes where the red and the blue colour represent maximum electron density and minimum electron density respectively. The yellow colour code, which lies precisely in between the red and blue zones, represents a neutral region. The ESP maps of sequential $H_2$ adsorption over $Si_4Li$, $Si_4Li_2$, and $Si_4Li_3$ binary clusters are depicted in Fig. 3.

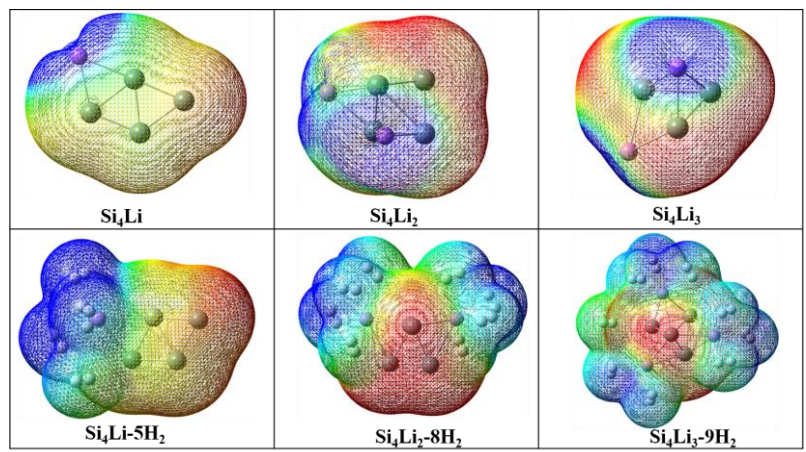

**Fig. 3**: ESP maps of bare hosts and their saturated $H_2$ adsorbed clusters.

The figure clearly shows that the colour code over the Li in the binary clusters is bluish, which means that this region is deficient in electron density. However, as sequential adsorption of $H_2$ molecules over these clusters proceeds to saturation, the dark bluish colour over Li fades, demonstrating electron density redistribution due to adsorption phenomena. Moreover, it also indicates polarization of $H_2$ molecules due to the dipole's electrostatic field, consisting of partially positively charged Li atoms and partially negative charged Si atoms.

*3.2.2 Topological analysis*

Topological studies have been carried out using Bader's Quantum Theory of Atoms In Molecules (QTAIM) to investigate the interaction between the Li sorption centers and the adsorbed $H_2$ molecules. At the bond critical point (BCP) of a bond between two atoms, the electron density distribution is mainly characterized by the electron density ($\rho$), Laplacian of electron density ($\nabla^2\rho$), and the total energy density ($\mathcal{H}$). Bader's analysis demonstrates that if $\nabla^2\rho > 0$, i.e., Laplacian of electron density is positive, it indicates electron density depletion near the bonding region and the interaction is considered a closed-shell type or non-covalent interaction.

Besides this, the positive value of total energy density also confirms non-covalent interaction. Since our interest is confined to studying the Li-$H_2$ interaction, so, in Table 2, we have only considered the parameters associated with electron density distribution at the Li-$H_2$ bond critical point. Table 2 shows that for all the hydrogenated systems, the Laplacian of electron density ($\nabla^2\rho$) and the total energy density ($\mathcal{H}$) for the bond critical point of Li-$H_2$ is positive indicating a weak non-covalent type of interaction between Li and $H_2$ molecules.

**Table 2**: Electron density ($\rho$), Laplacian of electron density ($\nabla^2\rho$), and total energy density ($\mathcal{H}$) at BCP of Li-$H_2$ for saturated hydrogenated clusters.

| Complex | $\rho$ (Li-$H_2$) (a.u) | $\nabla^2\rho$(Li $-$ $H_2$) | $\mathcal{H}_{BCP}$ (Li $-$ $H_2$) (a.u) |
|---|---|---|---|
| $Si_4Li$-$5H_2$ | 0.007285 | 0.0418 | 0.002197 |
| $Si_4Li_2$-$8H_2$ | 0.007571 | 0.0434 | 0.002176 |
| $Si_4Li_3$-$9H_2$ | 0.007319 | 0.0420 | 0.002182 |

## Conclusions

In conclusion, we have investigated the hydrogen storage capacities of $Si_4Li$, $Si_4Li_2$, and $Si_4Li_3$ binary clusters. The stability of hydrogenated clusters was checked and confirmed by global reactivity descriptors. Sequential adsorption of $H_2$ molecules over $Si_4Li$, $Si_4Li_2$ and $Si_4Li_3$ reveal that each of Li in the mentioned clusters can adsorb $5H_2$, $4H_2$ and $3H_2$ molecules resulting in a total gravimetric density of 7.8%, 11.3% and 12% respectively which is well above the target set by US-DOE (5.5% by 2020). It is also found that the adsorption energy of $H_2$ molecules lies in the range of 0.12eV-0.17eV, which gives a clear indication of the adsorption of $H_2$ molecules in a quasi-molecular fashion via Niu-Rao-Jena type of interaction. In addition, electrostatic potential maps demonstrate redistribution of electron density during sequential adsorption of $H_2$ molecules. Topological studies reveal that interaction between Li sorption centers and the adsorbed $H_2$ molecules is non-covalent type. Hence, our findings confirm that the studied binary clusters can be considered potential hydrogen storage materials.

## Acknowledgements

We gratefully acknowledge IIT(ISM) Dhanbad for providing research facilities. Authors also acknowledge the financial supports provided by Science and Engineering Research Board (SERB), DST, India for the extra –mural research grant (EMR 2014/000141).